\documentclass{article} 

\usepackage{algorithm}
\usepackage{algorithmicx}
\usepackage{algpseudocode}
\usepackage{booktabs}
\usepackage{array}
\usepackage{graphicx}
\usepackage{amsmath}
\usepackage{amsfonts}
\usepackage{amssymb}
\usepackage{multirow}
\usepackage{url}

\begin{document}

\title{Efficient and flexible simulation-based sample size determination for clinical trials with multiple design parameters}

\author{Duncan T. Wilson\textsuperscript{1} \and
	Rebecca E. A. Walwyn\textsuperscript{1} \and
	Richard Hooper\textsuperscript{2} \and
	Julia Brown\textsuperscript{1} \and
	Amanda J. Farrin\textsuperscript{1} }
\date{1 - Leeds Institute of Clinical Trials Research, University of Leeds, Leeds, UK \\ 2 - Centre for Primary Care \& Public Health, Queen Mary University of London, London, UK}

\maketitle

\begin{abstract}
Simulation offers a simple and flexible way to estimate the power of a clinical trial when analytic formulae are not available. The computational burden of using simulation has, however, restricted its application to only the simplest of sample size determination problems, minimising a single parameter (the overall sample size) subject to power being above a target level. We describe a general framework for solving simulation-based sample size determination problems with several design parameters over which to optimise and several conflicting criteria to be minimised. The method is based on an established global optimisation algorithm widely used in the design and analysis of computer experiments, using a non-parametric regression model as an approximation of the true underlying power function. The method is flexible, can be used for almost any problem for which power can be estimated using simulation, and can be implemented using existing statistical software packages. We illustrate its application to three increasingly complicated sample size determination problems involving complex clustering structures, co-primary endpoints, and small sample considerations.
\end{abstract}

\section{Introduction}\label{sec:intro}

The sample size of a clinical trial is typically minimised subject to the power of the trial being above a nominal level, often 80 or 90\%. For many sample size determination (SSD) problems, power can be calculated using a simple mathematical formula and the optimisation problem can be solved in a timely manner. When complexity in the trial design or the method of analysis mean such formulae are not readily available, we can estimate power using a Monte Carlo (MC) approximation~\cite{Arnold2011, Landau2013}. To do so, we simply simulate several hypothetical sets of trial data under the alternative hypothesis, analyse each of these, and calculate the proportion of analyses which reject the null hypothesis. The simplicity and flexibility of the simulation method has seen it used for a variety of statistical models and study designs, including problems involving hierarchical models~\cite{Feng1992, Hooper2013}, proportional hazards models~\cite{Schoenfeld2005}, logistic regression models~\cite{Grieve2016}, individual patient data meta-analyses~\cite{Sutton2007, Kontopantelis2016}, patient enrolment models~\cite{Fedorov2005}, stepped wedge designs~\cite{Baio2015, Hooper2016}, and cluster randomised crossover designs~\cite{Reich2012}. Although calculating MC estimates of power can be computationally demanding, these SSD problems remain feasible because, as optimisation problems, they are quite simple. In particular, optimisation takes place over a single parameter (the sample size), subject to a single constraint (power), and with respect to a single objective to be minimised (the sample size again).

SSD problems, particularly those found in trials of complex interventions, are not always this simple~\cite{Wilson2015}. There may be several parameters, each influencing the power of the trial, which need to be specified at the design stage. Several design parameters are common in, for example, trials with multilevel structures such as cluster randomised trials, where both the number of clusters and the number of participants in each cluster must be specified. Increasing the number of design parameters complicates the SSD problem by increasing the number of possible solutions to search. A second complication arises when there is more than one criterion we are interested in minimising, subject to the nominal power constraint. A cluster randomised trial will often have this property, as we would like to minimise both the total number of participants and the number of clusters. Given multiple conflicting objectives, there is no single `optimum' solution but rather a range of solutions which offer different degrees of trade-off between the objectives. Seeking a set of good solutions, rather than a single optimum, further adds to the difficulty of the SSD problem.

Complex SSD problems with several design parameters and several objectives could in theory be solved using  benchmark multi-objective optimisation algorithms such as NSGA-II~\cite{Deb2002}, robust implementations of which are freely available in statistical software such as R~\cite{Mersmann2014}. However, these so-called `greedy' algorithms typically assume that evaluating any proposed solution to the problem is a very fast process, and consequently evaluate many thousands of solutions during the search. If these algorithms were applied to problems where evaluating solutions required computing an MC estimate of power, they would take an infeasibly long time to converge. Thus, if we are to extend simulation-based trial design to complex SSD problems, we require a more general framework employing more efficient optimisation algorithms. 

%[RW - have these greedy algorithms been used in the previously cited software? A: No. In fact, they have not been used in at least some cases where they would have been very useful - some examples of multi-stage designs with an analytic power function, multiple objectives, people searching from `admissible` designs by using lots of different scalarised functions and doing single objective optimisation for each - very inefficient.]

Outwith the context of clinical trial design, a great deal of research has addressed optimisation problems where the evaluation of a solution is a computationally demanding, or \emph{expensive}, operation~\cite{Sacks1989,Santner2003}. One approach addresses the problem by substituting the expensive function with a mathematical approximation known as a \emph{surrogate model}. The surrogate model is then used to make predictions about the true function for different values of design parameters, with these predictions informing which point should be evaluated next. The information obtained from this evaluation is used to update the surrogate model, thus improving the predictions available at the next iteration. One class of surrogate model is Gaussian process (GP) regression. Also known as Kriging and having its roots in geostatistics~\cite{Krige1951}, GP models are spatial interpolators which are computationally tractable~\cite{Rasmussen2006} and can be fitted using robust and freely available software~\cite{Roustant2012}. A GP surrogate model provides not only a prediction of the true function value at any point, but also a measure of uncertainty in this prediction. This property is exploited by the benchmark Efficient Global Optimisation (EGO) algorithm~\cite{Jones2001}, allowing the next point in the search to be chosen in a way that formally balances the potential benefits of \emph{exploitation} (searching around areas already known to be promising) and \emph{exploration} (searching in areas of high uncertainty). Although EGO was originally proposed for unconstrained optimisation of expensive objective functions with deterministic output, various proposals have extended it to incorporate the expensive constraints~\cite{Sasena2002}, multiple objectives~\cite{Emmerich2011}, and stochastic outputs~\cite{Picheny2014} that feature in complex SSD problems.

In this paper we will explore how GP regression models and a variant of the EGO algorithm can be used to solve complex SSD problems. In contrast with many of the available methods and software for simulation-based SSD, which focus on specific application areas such as multilevel designs (MLPowSim~\cite{Browne2009}), IPD meta-analyses (ipdpower~\cite{Kontopantelis2016}) or stepped wedge design (SWSamp~\cite{Baio2015}), we take the same approach as that used in the SimSam package \cite{Hooper2013} and propose a more general framework which can be applied to a broad class of SSD problems following the Neyman-Pearson hypothesis testing formulation. Specifically, we require that the user can write a program which simulates the data generating process and analysis of the trial, returning a binary indicator denoting rejection or otherwise of the null hypothesis. This flexibility will not only mean simulation-based SSD can be used for a wide range of existing trial designs, but will also facilitate SSD for novel designs developed in the future and which cannot be anticipated now.

The remainder of the paper is structured as follows. Three motivating problems are described in Section~\ref{sec:examples}. In Section~\ref{sec:prelim} we provide the necessary background and notation regarding Monte Carlo estimation and multi-objective optimisation. In Section~\ref{sec:methods} we describe Gaussian process regression, the efficient global optimisation algorithm, and a framework for its application to sample size determination. We return to the examples in Section~\ref{sec:application}, illustrating how the method can be applied in practice. We conclude with a discussion of the implications and limitations of the proposed approach in Section~\ref{sec:discussion}.

\section{Motivating examples}\label{sec:examples}

`Pacing, graded Activity, and Cognitive behaviour therapy; a randomised Evaluation' (PACE)~\cite{White2007, White2011} was a randomised controlled trial comparing adaptive pacing therapy (APT), cognitive behavioural therapy (CBT), graded exercise therapy (GET) and specialist medical care (SMC) as secondary care treatments for patients with chronic fatigue syndrome (CFS/ME). The data had a complex multilevel structure, with three of the four arms including a therapy provided by different therapists (partially nested structure) and all arms including medical care from the same doctors (crossed structure), leading to the potential for treatment-related clustering. Since some participants receive treatment from both a therapist and a doctor, the relationship between participants and care providers is cross-classified. Moreover, two potentially correlated co-primary endpoints (fatigue and disability) were used. The original sample size calculation used a simple analytic formula for comparing proportions in two groups of equal size. As was typical at the time, it did not account for the impact of clustering or of simultaneously analysing two correlated endpoints. In this section we will describe three theoretical example SSD problems based around the PACE trial, increasing in complexity at each step. 

\subsection{Complex clustering}\label{sec:ex1}

\begin{figure}
\centering
\includegraphics[scale=0.7]{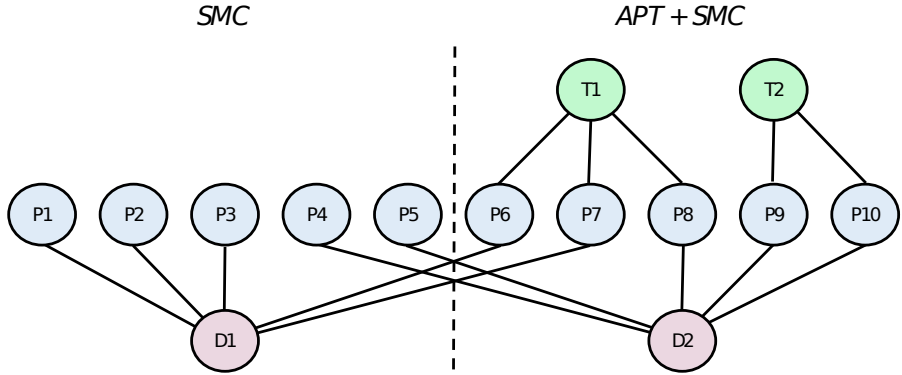}
\caption{Multilevel structure of the SMC and APT arms of our example, where therapists (T) are partially nested within interventions, doctors (D) are crossed with interventions, and patients (P) are cross-classified with therapists and doctors in the intervention arm and nested within doctors in the control arm.}
\label{fig:ex1_structure}
\end{figure}

For simplicity, we consider redesigning the PACE trial to detect a difference in the probability of participants responding with respect to fatigue between the APT and SMC arms. A participant is considered to have responded if they have a score of 3 or less (indicating normal fatigue) on the likert Chalder Fatigue Scale (CFS) \cite{Chalder1993}. As in the original design, we assume an equal number $n$ of participants will be recruited to each arm. In the intervention arm, $k$ therapists will deliver APT to participants, with each participants receiving treatment from a single therapist. We assume that the number of participants allocated to each therapist will vary with therapist. Specifically, we model the proportion of all participants in the APT arm allocated to a therapist using a Gamma distribution with shape parameter 1. Participants in both APT and SMC arms will receive specialist medical care from one of $2k$ doctors, with the proportion allocated to each doctor also following a gamma distribution with shape parameter 1. This leads to a multilevel data structure where therapists are partially nested within interventions, doctors are crossed with interventions, and patients are cross-classified with therapists and doctors in the intervention arm and nested within doctors in the control arm \cite{Walwyn2010}. This structure is illustrated in Figure~\ref{fig:ex1_structure}.

In this example, the primary analysis will be a logistic mixed effect model with a likelihood ratio test of the null hypothesis that the probability of response in the APT arm, $p_1$, is equal to that in the SMC arm, $p_0$. For simplicity, the model includes random intercepts in the linear predictor for doctor effects and a random coefficient for therapist effects. Further work is needed on the model that is recommended in this scenario but as the model we assume is for illustration purposes only, it could simply be updated once a recommended model becomes available. We assume that the test statistic follows a chi-squared distribution with 1 degree of freedom, and therefore that the type I error rate can be controlled at a nominal level of $\alpha^* = 0.05$ (two-sided). As model convergence may be an issue, we include this in our definition of power by not rejecting the null hypothesis when the model fails to converge. We require that the power at the alternative hypothesis $H_1: p_0=0.1, p_1=0.25$ be no less than 90\%. Subject to these constraints, we aim to find minimal values of $n$ and $k$, recognising that these two objectives will conflict with one another. 

Although sample size formulae for partially nested designs with binary outcomes are available~\cite{Roberts2015}, they do not extend to the structure in this example where participants are cross-classified and doctors are crossed with interventions. Including the complex process of model non-convergence into the definition of power further necessitates the use of simulation.

\subsection{Co-primary endpoints}\label{sec:ex2}

We extend the previous example to include a co-primary endpoint relating to disability, a binary response defined as 75 (out of 100) or more on the short form-36 physical function subscale \cite{McHorney1993}, where the mean score for the UK adult population is around 85. An in PACE, the endpoints will be analysed separately, each time fitting a logisitc mixed effect model and conducting a likelihood ratio test as described in Section~\ref{sec:ex1}. The results of the trial will be considered positive only if both of the analyses show a statistically significant difference, leading to reduced power under the alternative hypothesis of an effect on each endpoint in comparison to the univariate case of Section \ref{sec:ex1}\cite{Senn2007}. Correlation between the endpoints is expected at the participant level, and correlations between the random effects of therapists and doctors for different endpoints are also expected. This correlation will be modelled in the data generating process, but not in the analysis due to concerns about the feasibility of fitting a multivariate model. Such a mismatch between the data generating and analysis models has been noted as a clear motivation for simulation-based power calculations \cite{Landau2013}.

\subsection{Small sample pilot}\label{sec:ex3}

Finally, we consider how we might have designed a pilot trial prior to the definitive PACE trial to provide a preliminary test of potential efficacy. At this early stage, we would like 90\% power to detect a meaningful effect in \emph{either} the fatigue or disability endpoints. To enable a small trial to have such high power, we relax the type I error rate to 0.2 (one-sided) and change the primary endpoints from binary responses to the continuous scores on the CFS and SF-36. In the pilot setting we assume we have greater control over the numbers of participants allocated to therapists and to doctors, and so can maximise efficiency by balancing cluster sizes. We also now consider varying the number of doctors.  Our objectives are to minimise the total number of participants, the number of therapists, and the number of doctors. The small sample setting of a pilot trial implies the distributional assumptions underpinning type I error control may be violated, and so we simulate power under the null hypothesis and model this constraint in addition to power under the alternative. We include the nominal type I error rate used when testing the null hypothesis as a design variable, allowing an appropriate adjustment to be made as part of the larger optimisation process.
 
In terms of design parameters, we must choose the total sample size in the APT arm, denoted $n_1$; the number of APT therapists, $k$; the allocation ratio relating the total number of participants in each arm, $r = n_0/n_1$; the number of doctors, $j$; and the nominal type I error rate, $a$. Thus, in comparison with the preceding examples, the number of designs to be searched over is significantly larger. By requiring the simulation of power under the null and alternative hypotheses, the computational burden of simulation is doubled. By minimising three objectives simultaneously, a larger set of solutions will be required to enable the available trade-offs between them to be fully appreciated. 

\section{Background}\label{sec:prelim}

\subsection{Monte Carlo estimation}\label{sec:MC}

Monte Carlo methods can be used to numerically approximate expectations $\mathbb{E}[f(Z)]$ of real valued functions $f(Z)$ with respect to the probability distribution of $Z$. Given $N$ samples of $Z$, denoted $z_{i}$, $i=1,\ldots,N$, the MC estimate is
\begin{equation}
\mathbb{E}[f(Z)] \approx \frac{1}{N} \sum_{i=1}^{N} f(z_{i}).
\end{equation}
The estimate is unbiased for all $N$ and has variance equal to
\begin{equation}
\omega^{2} = Var \left[ \frac{1}{N} \sum_{i=1}^{N} f(z_{i}) \right] = \frac{1}{N} Var \left[ f(z_{i}) \right].
\end{equation}
The standard error of the MC estimate will therefore reduce at a rate of $1/\sqrt{N}$ as we increase $N$. When $N$ is large we can consider an MC estimate to be the true expectation plus a normally distributed error term with 0 mean and variance $\omega^{2}$, i.e.
\begin{equation}\label{eqn:MC_error}
\frac{1}{N} \sum_{i=1}^{N} f(z_{i}) = \mathbb{E}[f(Z)] + e \text{, where } e \sim N(0, \omega^{2}).
\end{equation}

In the context of simulation-based trial design, if $Z$ is the test statistic to be compared with an acceptance region $\Lambda$ then the probability of acceptance under hypothesis $H$ is $\mathbb{E}[I(Z \in \Lambda) \mid H]$, where $I(.)$ is the indicator function. An MC estimate of the power of a trial design under $H$ can therefore be obtained given $N$ test statistics $z_{1}, \ldots , z_{N}$ sampled under $H$. The steps required to simulate these statistics are described in~\cite{Landau2013}, and we briefly summarise them here:
\begin{enumerate}
\item Define the population model. This describes the underlying target population and should specify all population parameters and distributions under the hypothesis of interest.
\item Define the sampling strategy. This should specify the numbers of patients, clusters, or any other sampling units in the trial and how they will be drawn from the population.
\item Define the method of analysis. For hypothesis testing, this will include defining the form of the test statistic $Z$ and the acceptance region $\Lambda$.
\end{enumerate}
Given each of the above elements, pseudo-random number generators can be used to simulate the recruitment, randomisation and primary outcome measure of patients under the hypothesis of interest, from which a test statistic $z_{i}$ can be calculated. 

\subsection{Multi-objective optimisation}\label{sec:optimisation}

A solution to the SSD problem consists of a vector of design parameters $\mathbf{x}$, and the \emph{solution space} $\mathcal{X}$ is the set of all solutions. A simple SSD problem may have a 1-dimensional solution space, while more complex problems may have several dimensions. Elements of $\mathbf{x}$ may include parameters defining the sample size of the trial, the acceptance region to be used in the analysis, or any other design aspect over which we have control and which may influence the trial operating characteristics. For instance, in example \ref{sec:ex1} a solution $\mathbf{x} = (k, n)$ is defined by the number of participants in each arm ($n$) and the number of therapists in the intervention arm ($k$).

An \emph{objective function} $f(\mathbf{x})$ is a function $f : \mathcal{X} \rightarrow \mathbb{R}$ which we wish to minimise. In a multi-objective problem with $B$ objectives, we denote the vector of objective values as $\mathbf{y} = (f_{1}(\mathbf{x}), \ldots, f_{B}(\mathbf{x})) \in \mathbb{R}^{B}$. We will describe $\mathbb{R}^{B}$ as the \emph{objective space}. In our example \ref{sec:ex1}, we have two objectives: minimising the number of clusters $f_1(\mathbf{x}) = k$; and minimising the total number of participants, $f_2(\mathbf{x}) = 2n$.

A \emph{constraint function} $g(\mathbf{x})$ is a function $g : \mathcal{X} \rightarrow \mathbb{R}$ which must be less than or equal to 0 for the solution $\mathbf{x}$ to be considered feasible. For example, if type II error rate is denoted by $\beta(\mathbf{x})$ and the nominal type II error rate is set at $\beta^{*}$, a constraint function would be $g(\mathbf{x}) = \beta(\mathbf{x}) - \beta^{*}$. We denote $C$ constraint functions as $g_{j}(\mathbf{x}),~j=1,\ldots , C$. The general SSD problem can now be stated as
\begin{align}
\min_{\mathbf{x} \in \mathcal{X}} {~ f_{i}(\mathbf{x})}, ~ i = 1, \ldots , B \\
\text{subject to} ~ g_{j}(\mathbf{x}) \leq 0, ~ j = 1, \ldots , C.
\end{align}

\begin{figure}
\centering
\includegraphics[scale=0.8]{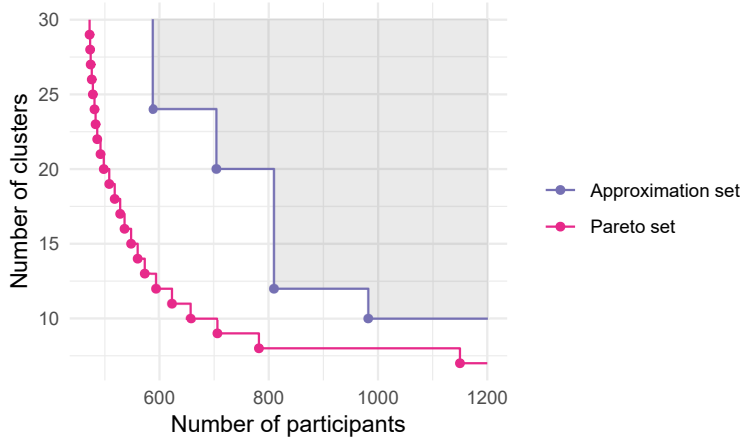}
\caption{Example Pareto front $\mathcal{X}_{p}$ and approximation set $\mathcal{A}$ for a cluster randomised trial design problem. The dominated hypervolume of the approximation set with respect to a reference point (cross) is the shaded area.}
\label{fig:fake_pareto}
\end{figure}

We denote by $\prec$ the relation of Pareto dominance, where a solution dominates another if it is at least as good in all respects, and better in at least one. Formally, $\mathbf{x}_{*} \prec \mathbf{x}$ if $f_{i}(\mathbf{x}_{*}) \leq f_{i}(\mathbf{x})$ for $i = 1, \ldots , B$, and $f_{j}(\mathbf{x}_{*}) < f_{j}(\mathbf{x})$ for some $j$. For instance, $\mathbf{x}_a = (n=100, k=10) \prec \mathbf{x}_b = (n=120, k=10)$ in example \ref{sec:ex1}, but $\mathbf{x}_a = (n=100, k=10) \nprec \mathbf{x}_c = (n=80, k=13)$. The \emph{Pareto set} is the set of non-dominated solutions $\mathcal{X}_{p} = \{\mathbf{x} \in \mathcal{X} \mid \nexists ~ \mathbf{x}_{*}  \in \mathcal{X} ~\text{s.t.}~ \mathbf{x}_{*} \prec \mathbf{x} \}$. An example Pareto set for example \ref{sec:ex1} is plotted in Figure~\ref{fig:fake_pareto}, illustrating the available trade-offs between the two objectives.

Multi-objective optimisation seeks to find a set of solutions that are close to the true Pareto set, with every member of the set non-dominated with respect to all other members. We will refer to these as approximation sets, denoted $\mathcal{A}$. That is, any set $\mathcal{A}$ such that $\mathcal{A} \in \mathcal{X}$ with $\forall~\mathbf{x} \in \mathcal{A}:~\nexists~\mathbf{x}_{*} \in \mathcal{A} : \mathbf{x}_{*} \prec \mathbf{x}$ is an approximation set \cite{Emmerich2011}. A set $\mathcal{A}$ is feasible if all constraints are satisfied by every member of $\mathcal{A}$. An example feasible approximation set for example \ref{sec:ex1}, plotted in Figure~\ref{fig:fake_pareto}, is given by the four $(2n, k)$ points
\begin{equation}
\mathcal{A} = \{ (589, 24), (705, 20), (810, 12), (982, 10) \}.
\end{equation}

To understand the similarity between any approximation set $\mathcal{A}$ and the ideal Pareto set, we measure its dominated hypervolume. This is the volume of the subspace dominated by solutions in $\mathcal{A}$ and bounded by a reference point $r$:
\begin{equation}
H(\mathcal{A}) = \text{Vol}(\{\mathbf{y} \in \mathbb{R}^{B} \mid \mathbf{y} \text{ is dominated by some } \mathbf{y}_{*} \in \mathcal{A} \text{ and } \mathbf{y} \prec r \}). 
\end{equation}
The largest possible hypervolume of any feasible approximation set $\mathcal{A}$ is achieved by the true Pareto set $\mathcal{X}_{p}$. We can therefore frame the multi-objective optimisation problem as finding the feasible approximation set $\mathcal{A}$ with largest hypervolume. Taking a reference point of $r = (1200, 30)$ (marked by the cross in Figure~\ref{fig:fake_pareto}), our example approximation set has a dominated hypervolume of 9202. This can be compared with that of the Pareto set, at 14501. We would expect the approximation set to converge to the Pareto set as the number of optimisation iterations increases.

\section{Simulation-based sample size determination}\label{sec:methods}

\subsection{Overview}

The proposed method is based on the Efficient Global Optimisation algorithm~\cite{Jones1998}. For clarity we will describe the algorithm in the context of an SSD problem with a single constraint function, denoted $g(\mathbf{x})$, which must be estimated using simulation. The more general case of several constraints will follow. The initial step is to select a number of potential solutions to the SSD problem $\mathcal{X}_{E} = (\mathbf{x}^{(1)}, \ldots , \mathbf{x}^{(E)})$ and evaluate the constraint function at each of these points, giving $\mathbf{y}_{E} = (g(\mathbf{x}^{(1)}), \ldots , g(\mathbf{x}^{(E)}))$. A Gaussian process regression model is then fitted to the data, relating the solutions $\mathcal{X}_{E}$ to the estimates $\mathbf{y}_{E}$ and providing an approximation of the constraint function $g$. The solution $\mathbf{x}_{*}$ which has the largest expected improvement $EI(\mathbf{x})$ according to the predictions of the GP model, is then found. This solution is evaluated to obtain $y_*$. This new data is then used to update the GP model, which is then used again to find the next solution to evaluate. The algorithm can be repeated until either the computational resources available have been exceeded, or until no further improvements are being obtained. The Algorithm is summarised in~\ref{alg:EGO} below.

\begin{algorithm}
\caption{Efficient Global Optimisation~\cite{Jones1998}}\label{alg:EGO}
\begin{algorithmic}[1]
\State Compute MC estimates $\mathbf{y}_{E} = (g(\mathbf{x}^{(1)}), \ldots , g(\mathbf{x}^{(E)}))$
\While{Computation budget not exhausted}
\State Regress $\mathbf{y_{E}}$ on $\mathcal{X}_{E} = (\mathbf{x}^{(1)}, \ldots , \mathbf{x}^{(E)})$
\State Find $\mathbf{x}_{*} = \arg\max EI(\mathbf{x})$
\State Compute MC estimate $y_{*} = g(\mathbf{x}_{*})$ and add to $\mathbf{y_{E}},~\mathcal{X}_{E}$
\State Update the computational budget
\EndWhile
\end{algorithmic}
\end{algorithm}

The process of computing MC estimates used in steps (1) and (5) has already been described in Section~\ref{sec:MC}. In what follows we will first consider step (3), describing Gaussian process regression models and outlining how they can be fitted and used to make predictions. The notion of expected improvement in step (4) will then be defined for the constrained multi-objective problems we are concerned with. Finally, we cover the remaining aspects of implementation.

\subsection{Gaussian process regression}\label{sec:GP}

Consider a set of points $\mathcal{X}_{E} = \{ \textbf{x}^{(1)}, \ldots , \textbf{x}^{(E)} \} \subset \mathcal{X}$ at which an expensive function $g$ will be estimated using the Monte Carlo method. Consider also some other point $\mathbf{x}_{*} \not\in \mathcal{X}_{E}$ where we are interested in making a prediction of $g(\mathbf{x}_{*})$. The value of $g$ at each point in $\{\mathcal{X}_{E}, \mathbf{x}_{*}\}$ is initially unknown, but can be modelled by a Gaussian process (GP). 

In using a GP we assume that our belief regarding the the values of $g$ can be represented by a multivariate normal distribution. Prior to computing any estimates of $g$, we assume that the mean function of this multivariate normal is equal to zero\footnote{This is not a restrictive assumption. After observing estimates of the function $g$ and updating the GP model to account for these, the mean function can take on non-zero values.}. We write the covariance matrix of the distribution as
\begin{equation}\label{eqn:cov_matrix}
\begin{pmatrix}
K(\mathcal{X}_{E}, \mathcal{X}_{E}) & K(\mathcal{X}_{E}, \mathbf{x}_{*}) \\
K(\mathbf{x}_{*}, \mathcal{X}_{E}) & K(\mathbf{x}_{*}, \mathbf{x}_{*})
\end{pmatrix},
\end{equation}
where $K(\mathcal{X}_{E}, \mathcal{X}_{E})$ is the $E \times E$ covariance matrix for the points $\mathcal{X}_{E}$, $\mathbf{k}_{*} = K(\mathcal{X}_{E}, \mathbf{x}_{*}) = K(\mathbf{x}_{*}, \mathcal{X}_{E})$ is the $E$-length vector of covariances between $\mathcal{X}_{E}$ and $\mathbf{x}_{*}$, and $K(\mathbf{x}_{*}, \mathbf{x}_{*})$ is the variance at $\mathbf{x}_{*}$.

Given this prior distribution, we compute the MC estimates $y^{(1)} , \ldots , y^{(E)}$ at each point in $\mathcal{X}_{E}$. From equation (\ref{eqn:MC_error}), $y^{(i)} = g(\textbf{x}^{(i)}) + e^{(i)}$ where $e^{(i)}$ is a zero-mean normally distributed error term with standard deviation $\omega^{(i)}$. We denote by $\Delta$ the $E \times E$ diagonal matrix where the $i$th entry is $[\omega^{(i)}]^{2}$. The distribution of $g(\mathbf{x}_{*})$ conditional on the observed $\mathbf{y}$ can be shown to be normal with mean $\mathbf{k}_{*}^\top(K + \Delta)^{-1}\mathbf{y}$ and variance $k(x_{*}, x_{*})-\mathbf{k}_{*}^\top(K + \Delta)^{-1}\mathbf{k}_{*}$~\cite{Rasmussen2006}. Thus, given a prior covariance matrix of the form (\ref{eqn:cov_matrix}) and some MC estimates of $g$ at the points $\mathcal{X}_{E}$, a conditional predictive distribution of $g(\mathbf{x}_{*})$ can be found. It is this distribution which will be used in the optimisation algorithm when deciding which solution should next be evaluated.

The predictive distributions are influenced by the prior covariance matrix (\ref{eqn:cov_matrix}). The matrix is populated using a covariance function (or \emph{kernel}), $k(\mathbf{x}, \mathbf{x}') : \mathcal{X} \times \mathcal{X} \rightarrow \mathbb{R}$. This function must be symmetric and positive definite for the covariance matrix to have the same properties. One such covariance function is the squared exponential, which has the form
\begin{equation}
k(\mathbf{x}, \mathbf{x}') = \sigma \exp \left( -\sum_{j=1}^{D} \frac{(x_{j} - x'_{j})^2}{\lambda_{j}^2} \right).
\end{equation}
By using covariance functions of this form we will obtain a Gaussian process which is infinitely differentiable over $\mathcal{X}$ and thus very smooth. This would appear to be a reasonable restriction to place upon the power functions we are interested in. In order to populate the covariance matrix we must choose values of the hyper-parameters $\boldsymbol{\theta} = (\sigma, \lambda_{1}, \ldots, \lambda_{D})$. We do this by numerically optimising the log marginal likelihood
\begin{equation}\label{eqn:loglik}
\log{p(\mathbf{y} \mid \mathcal{X}_{E}, \boldsymbol{\theta})} = -\frac{1}{2}\mathbf{y}^\top[K + \Delta]^{-1}\mathbf{y} - \frac{1}{2} \log{|K + \Delta|} - \frac{n}{2} \log{2\pi},
\end{equation}
considered as a function of $\boldsymbol{\theta}$~\cite{Rasmussen2006}. Fitting a GP model by maximum likelihood in this manner can be done using the function \texttt{km} in the R package DiceKriging, as illustrated in the appendix.

\begin{figure}
\centering
\includegraphics[scale=0.8]{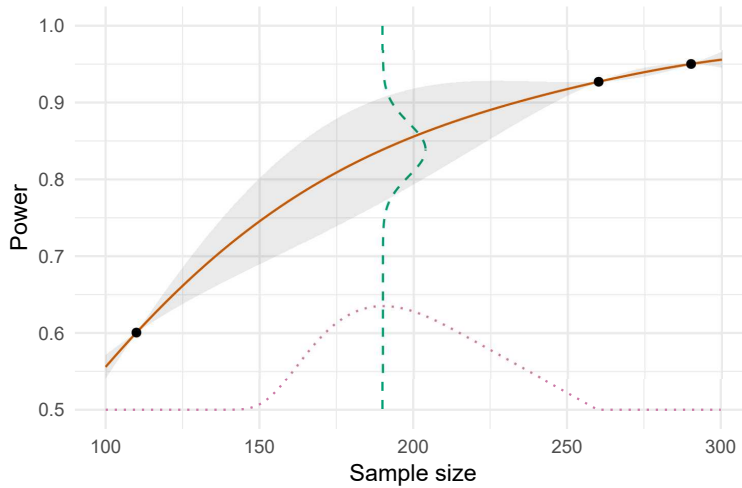}
\caption{A Gaussian process model of a power function over a one-dimensional sample size (solid line) based on three evaluations. Uncertainty is shown as the shaded area. Expected improvement (dotted line) is maximised at a sample size of 190, where the predicted power is normally distributed around a mean estimate of 0.84 (dashed line).}
\label{fig:GP_example}
\end{figure}

An illustration of a Gaussian process regression model of a power function in one dimension is given in Figure~\ref{fig:GP_example}. The power of three different choices of sample size have been calculated and a GP model fitted to the results. The figure illustrates how the uncertainty in the model predictions (shaded area) increases the further we are from a point which has been evaluated. The GP prediction of power at a sample size of $n = 190$, shown as a dashed line, is normally distributed with mean 0.84 and standard deviation 0.035.

\subsection{Expected improvement}\label{sec:EGO}

At any given point during the optimisation process we can obtain an approximation set $\mathcal{A}$ based on the set of solutions which have been evaluated up to that point. If a new point $\mathbf{x}_{*}$ is considered feasible, a new approximation set $\mathcal{A}_{*}$ will be identified. The improvement resulting from the evaluation of $\mathbf{x}_{*}$ is the difference in the dominated hypervolumes:
\begin{equation}
I = H(\mathcal{A}_{*}) - H(\mathcal{A}).
\end{equation}

Prior to evaluation, we do not know if the point $\mathbf{x}_{*}$ will be considered feasible. We therefore modify $I$ to account for the probability that $\mathbf{x}_{*}$ will be considered feasible after the MC estimates have been obtained. This probability can be estimated using the GP regression methodology described in Section~\ref{sec:GP}. A GP model of unknown constraint function $g$ will describe our current belief about the value of $g$ at $\mathbf{x}_{*}$ using a normal distribution with mean $m$ and variance $s^2$ $g(\mathbf{x}_{*}) \sim \mathcal{N}(m, s^{2})$, and we will consider the point $\mathbf{x}_{*}$ feasible if the upper $100 \times p$\% quantile of this distribution is below 0. We denote this quantile as
\begin{equation}\label{eqn:quant}
q(\mathbf{x}_{*}) = m + \Phi^{-1}(p)s,
\end{equation}
where $\Phi$ is the standard normal cumulative distribution. Following an evaluation of $\mathbf{x}_{*}$ the GP model will be updated and the quantile revised to $q_{+}(\mathbf{x}_{*})$. Before the evaluation the value of $q_{+}(\mathbf{x}_{*})$ is unknown, but it is shown in~\cite{Picheny2014} that its predictive distribution is $q_{+}(\mathbf{x}_{*}) \sim N(m_{+}, s_{+}^{2})$ where
\begin{align}
m_{+} &= m + \Phi^{-1}(p)\sqrt{\frac{\omega^{2}s^{2}}{\omega^{2} + s^{2}}} \\
s_{+}^{2} &= \frac{[s^{2}]^{2}}{\omega^{2} + s^{2}},
\end{align}
and $\omega$ is the MC error of the planned evaluation, estimated as $m(1-m)/N$ where $N$ is the number of MC samples to be used. The predictive distribution can then be used to calculate the probability that the point $\mathbf{x}_{*}$ will be considered feasible following its evaluation. Following~\cite{Sasena2002}, we multiply the theoretical improvement $I$ by this probability, thus penalising candidate solutions with a low chance of satisfying the constraint. This then gives us our expected improvement measure $EI$, where
\begin{equation}
\text{Expected Improvement } EI(\mathbf{x}_{*}) = [H(\mathcal{A}_{*}) - H(\mathcal{A})] \prod_{j=1}^{C} \Phi\left(\frac{-m_{j,+}}{s_{j,+}}\right).
\end{equation}
Note that we include a penalty term for all $j = 1, \ldots , C$ constraint functions. This maximisation problem is in itself complex, with a potentially large number of local maxima. We therefore use the particle swarm optimisation algorithm as implemented in the R package pso~\cite{Bendtsen2012}, designed to avoid becoming trapped in local maxima, to solve this sub-problem.

An illustration of expected improvement for a single-objective problem is given in Figure~\ref{fig:GP_example}. When choosing which sample size to evaluate next and aiming to find the lowest per-arm sample size with at least 80\% power, we balance the potential improvement over the best current solution (a sample size of 260) with the probability of constraint satisfaction. In this case, we would choose to evaluate the sample size of 190, estimated by the GP model to have a power of 84\%.

\subsection{Implementation}

To apply Algorithm~\ref{alg:EGO} in practice we must first choose the initial set of points to be evaluated, $\mathcal{X}_{E}$. One recommendation is to include 10 points for each dimension of the solution space, and to allocate between 30 and 50\% of the total computation budget to their evaluation\cite{Picheny2010}. To select the location of the points in $\mathcal{X}_{E}$ we use the space-filling Sobol sequence generated using the R package randtoolbox~\cite{Dutang2015}. The number of iterations and the number of MC samples $N$ used at each iteration must also be chosen. Given a total computational budget in terms of MC samples, the choice of these values should account for the fact that fitting GP regression models in R to more than around 800 points is currently infeasible~\cite{Chevalier2014}. 

As the algorithm depends on GP regression models, it can be helpful to assess the fit of these models. One approach is to regularly plot the predicted mean and standard deviation in one or two dimensions, centred at the last evaluated point. Poor model fit could be identified if the mean function is not, for example, strictly increasing as expected. We can also contrast the predicted function values with the obtained function values at each iteration, halting the algorithm if a large and unexpected discrepancy in these values is observed.

We have used R to implement the proposed framework, partly due to the availability of robust and efficient R packages for fitting Gaussian process models (DiceKriging~\cite{Roustant2012}) and for global optimisation (pso~\cite{Bendtsen2012}). Using R also provides flexibility in terms of the user-writen simulation routines by facilitating various complicated analysis procedures, e.g. multilevel modelling through lme4~\cite{Bates2015}. Our implementation works to a simple interface. The user must provide instances of two data frames. The first, \texttt{design\char`_space}, contains a row for each design parameter describing its name and its lower and upper bounds. The second, \texttt{constraints}, contains a row for each constraint function $g_{j}$. Each row should include a label for the constraint, the hypothesis it pertains to, a nominal power which should not be exceeded, and the confidence we require in the constraint being satisfied (i.e. the $p$ in Equation~\ref{eqn:quant}). Further, two functions are required. The first, \texttt{objectives}, takes as its argument a vector of design parameter values $\mathbf{x}$ and returns a vector of objective values $( f_{1}(\mathbf{x}), \ldots, f_{B}(\mathbf{x}) )$. The second, \texttt{sim\char`_trial}, takes as its arguments a vector of design parameter values $\mathbf{x}$ and a vector of parameter values defining the conditions under which we wish to simulate. The function should simulate the necessary data generation and analysis and return a boolean indicator of the rejection of the null hypothesis, or, more generally, of declaring `success'. Given these components, the example R code in the appendix can be modified to solve the problem at hand.

\section{Application to the examples}\label{sec:application}

In this section we describe the data generating models used to simulate trial data for each of our examples and the methods used in their analyses. Full details, including all the programs used to generate the results presented, are given in the appendix.

\subsection{Complex clustering}

We model the binary response of the $i$th participant using a latent variable representation. Specifically, we suppose that underlying the binary response $y_i$ there is a continuous latent variable $y_i^*$ such that
$$
y_i =
\begin{cases}
1 \text{ if } y_i^* \geq 0\\
0 \text{ if } y_i^* < 0.
\end{cases}
$$
As before, we define our model in terms of the $y_i^*$. Using the multilevel model notation of~\cite{Goldstein2003}:
\begin{align}\label{eqn:ex1_model}
y_i^* &= \beta_{0} + \beta_1 t_i + u_{therapist(i)}^{(2)}t_{i} + v_{doctor(i)}^{(2)} + e_i \\
u_{therapist(i)}^{(2)} & \sim N(0, \sigma_T^2) \\
v_{doctor(i)}^{(2)} & \sim N(0, \sigma_D^2),
\end{align}
where $e_i$ is a level 1 residual with mean zero and variance $\sigma_W^2$. Assuming $e_i$ follows a logistic distribution with $\sigma_W^2 = 3.29$ leads to a random intercept logistic model.

For the purposes of power calculations we must make some assumptions about the various nuisance parameter values. We set the 2nd level variance components to $\sigma_{T}^{2} = 0.19, \sigma_{D}^{2} = 0.37$ in order to give a variance partition coefficient of $\sigma_D^2 / (\sigma^2_D + \sigma^2_W) = 0.1$ in the control arm, a typical value in this setting. Similarly, the variance partition coefficient for between-therapist variance is then $\sigma^2_T/(\sigma^2_T + \sigma^2_D + \sigma^2_W) = 0.05$, and for between-doctor variation, $\sigma^2_D/(\sigma^2_T + \sigma^2_D + \sigma^2_W) = 0.095$. Recall that we want to simulate the power of the trial under the alternative hypothesis $H_1: p_0=0.1, p_1=0.25$. We can translate these probabilities into corresponding values for the coefficients in our model, giving $H_1: \beta_0 = \log(p_0/(1-p_0)) = -2.20, \beta_1 = \log(p_1/(1-p_1)) - \beta_0 = 1.10$.

The design parameters are the number of participants in each arm $n$, the number of therapists $k$ in the APT arm, and the number of doctors $j$ delivering specialist medical care across both arms. For simplicity and ease of illustration we will fix $j = 2k$. When searching over the remaining design parameters $n$ and $k$ we will initially consider $n \in [100, 500]$ and $k \in [3, 30]$, noting that these can be easily revised if the initial evaluations indicate larger values are required to achieve nominal power. We wish to minimise both the total number of patients $f_{1}(\mathbf{x}) = 2n$ and the total number of care providers $f_{2}(\mathbf{x}) = 3k$. The only constraint we must satisfy is that the type II error rate $\beta(\mathbf{x})$ under the alternative hypothesis is no more than $\beta^{*} = 0.1$. This gives the constraint function $g_{1}(\mathbf{x}) = \beta(\mathbf{x}) - 0.1$.

The original PACE sample size calculation did not account for clustering and, using simple analytic formulae for power of test of proportions, arrived at $n = 135$ per arm (before inflating for attrition) to achieve 90\% power. Simulating the actual power obtained from $n=135$ under our proposed model, with $k = 10$ therapists and $j = 20$ doctors, gave an MC estimate of 0.69 power (95\% confidence interval 0.66 to 0.72). Fitting two multilevel models for each sample led to a significant computational burden, needing over 5.3 minutes to generate the $N=1000$ samples required for this estimate. Thus, there is a need to search for an appropriate sample size using simulation, but a practical limit on the number of designs which we can evaluate in a timely manner. 

\begin{figure}
\centering
\includegraphics[scale=0.8]{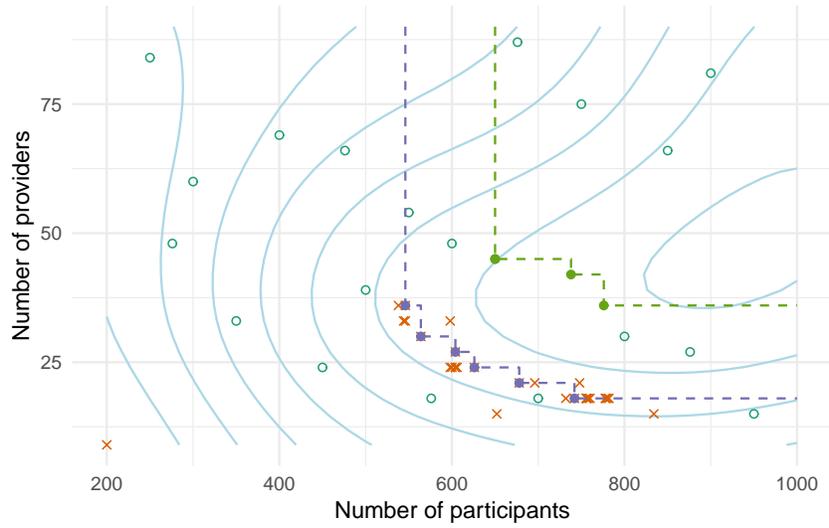}
\caption{Objective values of solutions in the initial set $\mathcal{X}_{E}$ (open circles), subsequent iterations of the algorithm (crosses), and those in the final approximation of the Pareto set (filled squares). The approximation set obtained using a fixed design is also shown (filled circles). Contours represent the mean function of the final GP model.}
\label{fig:ex1_single_run}
\end{figure}

We initialised the optimisation algorithm by generating a Sobol sequence of size 20 and computing MC estimates of power for each point using $N = 100$ samples. Following this, 30 iterations of the algorithm were applied, with $N = 100$ samples used at each iteration. We chose these optimisation parameters to ensure a solution could be found quickly, noting that further iterations can easily be added if solutions of a higher quality are sought. In Figure~\ref{fig:ex1_single_run} we plot the 50 evaluated solutions, distinguishing between those in the initial design $\mathcal{X}_{E}$, those which were subsequently evaluated during the iterative phase of the algorithm, and those which together form the final approximation set. The contours of the mean function of the final GP model are also shown. 

For comparison, we also plot the approximation set obtained using a similar procedure as that implemented in MLPowSim, software designed for simulation-based SSD for problems with multilevel data. Specifically, we take a Sobol sequence of size 50 and estimated the type II error at each of these points using $N = 100$ MC samples. For each point a 95\% confidence interval based on the MC error was calculated, and any points where the interval was not entirely below the nominal value of 0.1 were discarded. Of those that remained, any dominated solutions were discarded. The remaining two solutions are plotted in Figure~\ref{fig:ex1_single_run}. The proposed method has led to solutions of higher quality which collectively dominate those produced by the simpler method, with lower numbers of participants, providers, or both.

\begin{figure}
\centering
\includegraphics[scale=0.8]{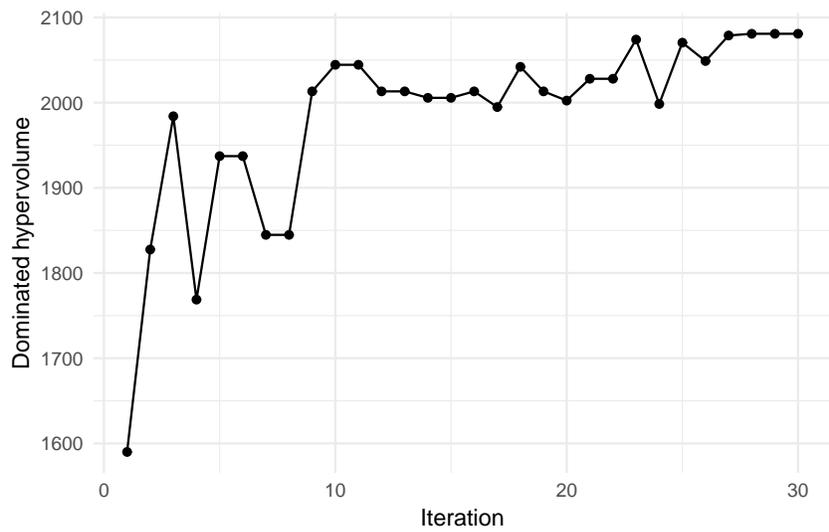}
\caption{Quality of the approximation set obtained as the algorithm proceeds, where higher dominated hypervolume reflects higher quality.}
\label{fig:ex1_traj}
\end{figure}

At the $i$th iteration of the algorithm we calculated the dominated volume $H(\mathcal{A}_{i})$, plotted in Figure~\ref{fig:ex1_traj}. In this instance the algorithm appears to successfully improve the quality of the approximation set as the number of iterations increases, with the rate of improvement decreasing over time. The total running time was 47 minutes. Note that $H(\mathcal{A}_{i})$ is not strictly increasing. This is because the evaluation of a new solution can lead to revised estimates of other solutions which were in the approximation set, such that they are then considered infeasible and removed from the set. 

The solutions which form the final approximation set are detailed in Table~\ref{tab:ex1_single_run}. As few as 6 therapists and 12 doctors can lead to a sufficiently powered trial, although 742 participants in total are required in this configuration. On the other hand, if minimising the total sample size is deemed more important than minimising the number of therapists and doctors, we see that as few as 546 participants are necessary providing 12 therapists and 24 doctors are included. The approximation set contains 6 solutions in total, providing a reasonable set of options from which the solution best representing the priorities of the decision makers can be selected.

\begin{table}
\small\sf\centering
\caption{Approximation set after 30 iterations for Example 1. Solutions are defined by their total sample size $2n$, number of therapists $k$, and number of doctors $j$. Type II error rate $\beta$ is constrained to be below 0.1, while the total sample size and number of providers are to be minimised.}
% latex table generated in R 3.4.3 by xtable 1.8-3 package
% Mon Apr 01 09:21:31 2019
\begin{tabular}{rrrll}
  \toprule
$2n$ & $k$ & $j$ & $\beta$ (s.e.), $N = 10^2$ & $\beta$ (s.e.), $N = 50^4$ \\ 
  \midrule
742 & 6 & 12 & 0.11 (0.031) & 0.093 (0.003) \\ 
  678 & 7 & 14 & 0.1 (0.03) & 0.081 (0.003) \\ 
  626 & 8 & 16 & 0.11 (0.031) & 0.087 (0.003) \\ 
  604 & 9 & 18 & 0.09 (0.029) & 0.077 (0.003) \\ 
  564 & 10 & 20 & 0.06 (0.024) & 0.081 (0.003) \\ 
  546 & 12 & 24 & 0.09 (0.029) & 0.084 (0.003) \\ 
   \bottomrule
\end{tabular}

\label{tab:ex1_single_run}
\end{table}

As can be seen from Table~\ref{tab:ex1_single_run}, approximate upper 95\% confidence limits based on the initial $N = 10^2$ MC estimates of power often exceed the corresponding nominal bound of 0.1. For example, the solution described in the first row would have an approximate upper 95\% two-sided confidence interval of $(0.049, 0.171)$. To verify the actual type II error we computed a more precise MC estimates  using $N = 50^4$ samples, which gave an estimate of 0.093 and a two-sided interval of $(0.087, 0.099)$. Similar results are seen for the remaining solutions in the approximation set, as shown in Table \ref{tab:ex1_single_run}. This demonstrates the GP's ability to share information of MC estimates computed at several points to increase the precision at each of them.

%However, the final GP model gives an estimate of 0.081 at this point, with a predicted interval of $(0.063, 0.099)$. 

\subsection{Co-primary endpoints}

For our second example we consider a second co-primary binary responder endpoint. We use the same latent variable representation as described in the preceding section to model the fatigue response $y_i^F$ and disability response $y_i^D$ of the $i$th participant. Correlation between these two endpoints is modelled by simulating bivariate residuals $(e_i^F, e_i^D)$ from a joint logistic distribution with correlation $\rho_W$ and marginal variances $\sigma_e^2 = 3.29$ as before. We also allow for correlation between the random effects associated with each therapist and doctor. These are now simulated according to the bivariate normal distributions
$$
(u_{therapist(i)}^F, u_{therapist(i)}^D) \sim N\left( (0,0)^T, \quad
\begin{pmatrix} 
\sigma_T^2 & \rho_T \sigma_T^2 \\
\rho_T \sigma_T^2  & \sigma_T^2 
\end{pmatrix}
\quad \right)
$$

$$
(v_{doctor(i)}^F, v_{doctor(i)}^D) \sim N\left( (0,0)^T, \quad
\begin{pmatrix} 
\sigma_D^2  & \rho_D \sigma_D^2  \\
\rho_D \sigma_D^2  & \sigma_D^2 
\end{pmatrix}
\quad \right)
$$
We set all correlations equal at $\rho_W = \rho_T = \rho_D = \rho_D = 0.9$, reflecting a situation where both a patient's responses and the individual therapist and doctor effects and very similar for both the fatigue and disability outcomes. 

\begin{figure}
\centering
\includegraphics[scale=0.8]{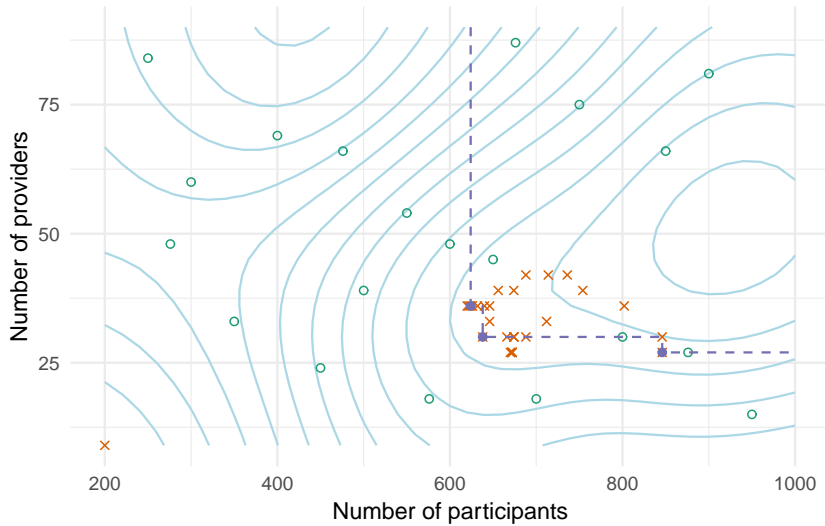}
\caption{Objective values of solutions in the initial set $\mathcal{X}_{E}$ (open circles), subsequent iterations of the algorithm (crosses), and those in the final approximation of the Pareto set (filled squares).}
\label{fig:ex2_single_run}
\end{figure}

The algorithm was applied using the same settings as before, with an initial design of 20 points followed by 30 iterations, and each evaluation using $N = 100$ MC samples. The run time in this case was 96 minutes, roughly double that of the previous example due to each simulation fitting twice as many models. The resulting approximation set is plotted in Figure~\ref{fig:ex2_single_run}. Again, the contours represent the mean function of the final GP model. 

Table \ref{tab:ex2_single_run} provides full details of the obtained approximation set together with their initial MC estimates of type II error rate using $N = 10^2$ MC samples, and further estimates using $N = 50^4$ MC samples. Approximate 95\% confidence intervals around the more precise estimates all either include the nominal constraint of 0.1, or lie entirely below it. 

\begin{table}
\small\sf\centering
\caption{Approximation set after 30 iterations for Example 2. Solutions are defined by their total sample size $2n$, number of therapists $k$, and number of doctors $j$. Type II error rate $\beta$ is constrained to be below 0.1, while the total sample size and number of providers are to be minimised.}
% latex table generated in R 3.4.3 by xtable 1.8-3 package
% Mon Apr 01 09:21:43 2019
\begin{tabular}{rrrll}
  \toprule
$2n$ & $k$ & $j$ & $\beta$ (s.e.), $N = 10^2$ & $\beta$ (s.e.), $N = 50^4$ \\ 
  \midrule
846 & 9 & 18 & 0.12 (0.033) & 0.069 (0.003) \\ 
  638 & 10 & 20 & 0.09 (0.029) & 0.104 (0.003) \\ 
  624 & 12 & 24 & 0.09 (0.029) & 0.096 (0.003) \\ 
   \bottomrule
\end{tabular}

\label{tab:ex2_single_run}
\end{table}

\subsection{Small sample pilot}

For our final example, recall that we have two continuous co-primary endpoints. For notational simplicity we use model (\ref{eqn:ex1_model}) but now consider the $y_i^*$ to be the actual observed continuous response, as opposed to a latent variable. We now assume the individual-level residual term $e_i$ is normally distributed but with the same variance as before, thus maintaining the variance partition coefficients at the same levels. The alternative hypothesis remains $H_1: \beta_1 = 1.10$. Note that this corresponds to a treatment effect standardised with respect to the total standard deviation in the APT arm of $1.10/\sqrt(0.19+0.37+3.29) = 0.56$.

Our design parameters (together with the ranges considered) are the total sample size in the APT arm, denoted $n_1$ (50 to 100); the number of APT therapists, $k$ (2 to 10); the allocation ratio relating the total number of participants in each arm, $r = n_0/n_1$ (0.5 to 1.5); the number of doctors, $j$ (3 to 20); and the nominal type I error rate to be used in the hypothesis tests, $a$ (0.05 to 0.2). The three objective functions to be minimised are $f_{1}(\mathbf{x}) = n_1 + rn_1$, $f_{2}(\mathbf{x}) = k$ and $f_{3}(\mathbf{x}) = j$. The two constraints to be satisfied are $g_{1}(\mathbf{x}) = \beta(\mathbf{x}) - 0.1$ and $g_{2}(\mathbf{x}) = \alpha(\mathbf{x}) - 0.2$.

\begin{figure}
\centering
\includegraphics[scale=0.8]{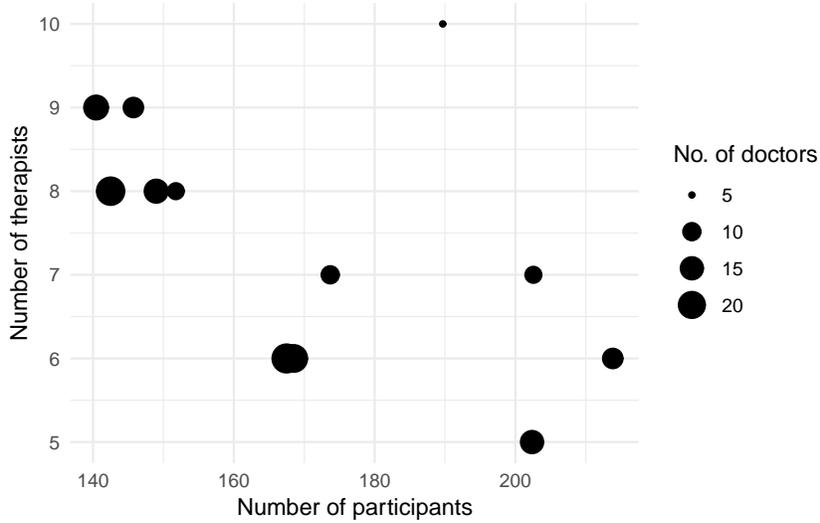}
\caption{Objective values of the approximation set obtained following 50 iterations of the algorithm for Example 3.}
\label{fig:ex3_single_run}
\end{figure}

Given the increase in dimensions of the solution space, we use an initial Sobol sequence design of 50 solutions. As before, we use 100 MC samples for each evaluation. After 50 iterations of the algorithm, an approximation set containing 15 solutions was obtained. The algorithm took 2 hours and 36 minutes to run. The objective values of these solutions are illustrated in Figure \ref{fig:ex3_single_run}, with full details provided in Table \ref{tab:ex3_single_run}. The total number of participants ranged from 140 to 214; of therapists, from 5 to 10; and of doctors, from 5 to 23. Type I error rates ranged from 0.09 to 0.14, all some way below the actual constraint value of 0.2. We calculated precise MC estimates (using $N = 50^4$ samples) of both type I and II error rates for each solution in the approximation set. As shown in Table \ref{tab:ex3_single_run}, type II error rates all appear to be around or slightly below the constraint of 0.1. Type I error rates, in contrast, are in some cases significantly below the constraint of 0.2. This suggests there is some potential for improvement in the approximation set by applying further iterations of the algorithm.

\begin{table}
\small\sf\centering
\caption{Approximation set after 50 iterations for Example 3. Solutions are defined by the number of participants in the APT arm $n_1$, the total number of participants across both arms $n$, the number of therapists $k$, the number of doctors $j$, and the nominal type I error rates $a$. Both Type I and II error rates are estimated using simulation using $N$ samples, and are constrained at 0.2 and 0.1 respectively.}
% latex table generated in R 3.4.3 by xtable 1.8-3 package
% Mon Apr 01 09:29:53 2019
\begin{tabular}{rrrrrllcll}
  \toprule
 & & & & & \multicolumn{2}{c}{$N = 10^2$} & \phantom{a} & \multicolumn{2}{c}{$N = 50^4$} \\ \cmidrule{6-7} \cmidrule{9-10} 
$n_1$ & $n$ & $k$ & $j$ & $a$ & $\beta$ (s.e.) & $\alpha$ (s.e.) && $\beta$ (s.e.) & $\alpha$ (s.e.) \\ 
  \midrule
94 & 202 & 5 & 15 & 0.11 & 0.07 (0.026) & 0.12 (0.033) && 0.088 (0.003) & 0.17 (0.004) \\ 
  94 & 169 & 6 & 21 & 0.12 & 0.11 (0.031) & 0.13 (0.034) && 0.081 (0.003) & 0.179 (0.004) \\ 
  94 & 214 & 6 & 12 & 0.11 & 0.05 (0.022) & 0.21 (0.041) && 0.09 (0.003) & 0.157 (0.004) \\ \vspace{2mm}
  84 & 167 & 6 & 23 & 0.10 & 0.09 (0.029) & 0.12 (0.033) && 0.103 (0.003) & 0.147 (0.004) \\ 
  79 & 174 & 7 & 10 & 0.12 & 0.17 (0.038) & 0.13 (0.034) && 0.086 (0.003) & 0.171 (0.004) \\ 
  95 & 203 & 7 & 9 & 0.09 & 0.1 (0.03) & 0.16 (0.037) && 0.085 (0.003) & 0.134 (0.003) \\ 
  75 & 152 & 8 & 9 & 0.13 & 0.1 (0.03) & 0.12 (0.033) && 0.083 (0.003) & 0.175 (0.004) \\ \vspace{2mm}
  78 & 149 & 8 & 16 & 0.12 & 0.08 (0.027) & 0.21 (0.041) && 0.088 (0.003) & 0.171 (0.004) \\  
  76 & 142 & 8 & 22 & 0.12 & 0.07 (0.026) & 0.17 (0.038) && 0.098 (0.003) & 0.156 (0.004) \\ 
  80 & 140 & 9 & 17 & 0.13 & 0.04 (0.02) & 0.2 (0.04) && 0.084 (0.003) & 0.176 (0.004) \\ 
  81 & 146 & 9 & 12 & 0.14 & 0.14 (0.035) & 0.18 (0.039) && 0.079 (0.003) & 0.181 (0.004) \\ 
  97 & 190 & 10 & 5 & 0.14 & 0.07 (0.026) & 0.27 (0.045) && 0.072 (0.003) & 0.178 (0.004) \\ 
   \bottomrule
\end{tabular}

\label{tab:ex3_single_run}
\end{table}

\section{Discussion}\label{sec:discussion}

Although simulation is often required for clinical trial sample size determination, related methodology has typically assumed that there is only one parameter which we are able to adjust (the sample size); that there is only one operating characteristic which must be estimated using simulation (the power of the trial); and that our goal is to minimise only one criterion (the sample size again)~\cite{Landau2013, Hooper2013}. In this paper we have described a flexible approach to simulation-based SSD which can be used for more general multi-parameter problems. The method draws on established global optimisation algorithms which use statistical `surrogate` models to solve design problems where there are several parameters to be chosen, several objectives to minimise, and several constraints to satisfy. We have illustrated how such problems arise in clinical trials of complex interventions.

The general optimisation framework we have suggested recognises that in many complex trials we are interested in minimising more than one quantity subject to constraints on operating characteristics. Problems of this sort are common in multilevel trial design~\cite{Hemming2017}, but are typically approached by first reducing the multiple objectives down to a single objective. For example, in the design of a cluster randomised trial it is common to fix the number of participants per cluster and minimise the number of clusters~\cite{Donner2000}, or vice-versa~\cite{Hemming2011, Eldridge2015}. Alternatively, a function which specifies the cost of sampling at the cluster and the patient level could be specified~\cite{Hox2002}, and the overall cost minimised~\cite{Snijders1993}. The latter approach has been suggested for both two-level~\cite{Raudenbush2000} and three-level hierarchical trial designs~\cite{Breukelen2012, Teerenstra2008}. However, the \emph{a priori} specification of such a cost function may not always be feasible, particularly when several stakeholders are involved \cite{Joseph1997a}. The Pareto optimisation framework we have described leads to a more computationally challenging optimisation problem, but produces a set of good solutions enabling the available trade-offs between objectives to be seen and selected from. 

As noted in Section~\ref{sec:intro}, related work in simulation-based design methodology has often focussed on a specific area of application. One advantage that brings is the relative ease with which the software can be used to solve a new problem within the same area. In contrast, our approach requires that the user provides a program which simulates the data generation and analysis of their proposed trial design. Although some have argued that this requirement may be prohibitive in practice~\cite{Kontopantelis2016}, it allows the user to solve their specific problem rather than some related version of it. Moreover, prior to addressing the sample size issue, modelling and simulation can help inform many other aspects of trial design, such as the patient population or the choice of endpoint~\cite{Smith2010}. One way to assist users in writing their own simulations is to share example programs for a range of problems, providing a starting point for the development of a new program. We have provided some examples in the appendix.

When submitting a proposed design for approval by a funding body it is important that the sample size calculation is transparent and replicable. This may be achieved in the context of simulation-based SSD by supplying the simulation program as part of the application~\cite{Hooper2013}. Given this, any reviewer should be able to re-calculate the operating characteristics of the proposed design. However, a greater challenge for the reviewer is understanding the program and ensuring it is an accurate representation of the model in question. This requirement has partly motivated our use of R. Although significantly slower than a compiled language such as C++, it has been argued that software written in R is more transparent~\cite{Smith2010}. Validation will be further facilitated if a simulation protocol of the sort described in~\cite{Burton2006} is provided alongside the code. Future work could develop an interface for alternative statistical software such as Stata or SAS, allowing a simulation program to be written in them and connect with the R implementation of the optimisation algorithm.

We have followed the conventional approach to clinical trial design whereby constraints on operating characteristics are set and then a constrained optimisation problem is solved. In practice the constraints are not fixed in advance, but adjusted iteratively in response to the design requirements they produce. For example, an initial nominal power of 90\% may require an infeasibly large sample size, leading to a revision down to 80\%. Such an iterative procedure will increase an already substantial computational burden for simulation-based design. However, note that a change to a constraint does not mean starting the process again, since any previous MC estimates can still be used when fitting the GP model(s). The sequential nature of the optimisation algorithm suggests that an interactive routine could be developed, where the user pauses the algorithm in response to the sample size requirements which are being observed, adjusts the constraints, and then continues with the optimisation.

The examples have demonstrated that the time required to solve a sample size determination problem can be significant, of the order of hours. Given that the majority of computational effort is expended generating MC samples when evaluating solutions, it is important that these simulation programs are as efficient as possible. We recommend making use of code profilers such as R's `Rprof' to identify the parts of the program that are consuming the most resources. Further efficiencies could potentially be gained by using more sophisticated methods for surrogate modelling and efficient optimisation. For example, when the modelled function can be assumed monotonic, this information can be incorporated into the surrogate modelling process~\cite{Emmerich2011}.

Numerous extensions to the proposed approach can be considered. One argument for simulation-based design is the ease with which sensitivity to model assumptions, such as the value of nuisance parameters, can be assessed~\cite{Landau2013}. Future work could consider how a systematic assessment of sensitivity to nuisance parameters could be conducted, given a proposed trial design. Such investigations fall under the heading of \emph{uncertainty quantification} and can be carried out using GP regression and associated techniques~\cite{Kennedy2001}. A further extension could consider Bayesian approaches to trial design, including hybrid Bayesian-frequentist assurances~\cite{OHagan2005}, fully Bayesian measures such as average coverage criterion~\cite{Cao2009}, and decision-theoretic methods~\cite{Oakley2010}. Aside from very simple cases involving only conjugate analyses, evaluating these Bayesian criteria will generally require simulation~\cite{OHagan2005} and so optimal design may benefit from the efficient methods discussed here. Complex SSD problems are also common in the area of adaptive designs, which can aim to minimise the expected sample size under several different hypotheses and over a number of stopping rule parameters~\cite{Wason2012}.  Extending the proposed methods to such problems would require using surrogate models to approximate the objective functions, as opposed to only the constraints.

In conclusion, efficient optimisation algorithms based on surrogate models of expensive operating characteristic functions can be used to solve complex clinical trial sample size determination problems. By using these methods we can avoid making unrealistic simplifying assumptions at the trial design stage, both in terms of the statistical model underlying the trial and of the nature of the design problem.

\subsection*{Funding}

This work was supported by the Medical Research Council [grant number MR/N015444/1].

\subsection*{Data availability statement}

All simulated data used in this manuscript, together with the code used to generate it, is available at \url{https://github.com/DTWilson/Bayes_opt_SSD}.

\bibliographystyle{unsrt}
\bibliography{C:/Users/meddwilb/Documents/Literature/Databases/DTWrefs}

\begin{thebibliography}{10}

\bibitem{Arnold2011}
Benjamin~F Arnold, Daniel~R Hogan, John~M Colford, and Alan~E Hubbard.
\newblock Simulation methods to estimate design power: an overview for applied
  research.
\newblock {\em {BMC} Med Res Methodol}, 11(1), 6 2011.

\bibitem{Landau2013}
Sabine Landau and Daniel Stahl.
\newblock Sample size and power calculations for medical studies by simulation
  when closed form expressions are not available.
\newblock {\em Statistical Methods in Medical Research}, 22(3):324--345, 2013.

\bibitem{Feng1992}
Ziding Feng and James~E. Grizzle.
\newblock Correlated binomial variates: Properties of estimator of intraclass
  correlation and its effect on sample size calculation.
\newblock {\em Statistics in Medicine}, 11(12):1607--1614, 1992.

\bibitem{Hooper2013}
Richard Hooper.
\newblock Versatile sample-size calculation using simulation.
\newblock {\em The STATA Journal}, 13(1):21--38, 2013.

\bibitem{Schoenfeld2005}
David~A. Schoenfeld and Michael Borenstein.
\newblock Calculating the power or sample size for the logistic and
  proportional hazards models.
\newblock {\em Journal of Statistical Computation and Simulation},
  75(10):771--785, oct 2005.

\bibitem{Grieve2016}
Andrew~P. Grieve and Shah-Jalal Sarker.
\newblock Simulation-based sample-sizing and power calculations in logistic
  regression with partial prior information.
\newblock {\em Pharmaceutical Statistics}, 15(6):507--516, sep 2016.

\bibitem{Sutton2007}
Alexander~J. Sutton, Nicola~J. Cooper, David~R. Jones, Paul~C. Lambert, John~R.
  Thompson, and Keith~R. Abrams.
\newblock Evidence-based sample size calculations based upon updated
  meta-analysis.
\newblock {\em Statistics in Medicine}, 26(12):2479--2500, 2007.

\bibitem{Kontopantelis2016}
Evangelos Kontopantelis, David~A Springate, Rosa Parisi, and David Reeves.
\newblock Simulation-based power calculations for mixed effects modeling:
  ipdpower in stata.
\newblock {\em Journal of Statistical Software}, 74(12), 2016.

\bibitem{Fedorov2005}
Valerii Fedorov and Byron Jones.
\newblock The design of multicentre trials.
\newblock {\em Statistical Methods in Medical Research}, 14(3):205--248, 2005.
\newblock PMID: 15969302.

\bibitem{Baio2015}
Gianluca Baio, Andrew Copas, Gareth Ambler, James Hargreaves, Emma Beard, and
  Rumana~Z Omar.
\newblock Sample size calculation for a stepped wedge trial.
\newblock {\em Trials}, 16(1), aug 2015.

\bibitem{Hooper2016}
Richard Hooper, Steven Teerenstra, Esther de~Hoop, and Sandra Eldridge.
\newblock Sample size calculation for stepped wedge and other longitudinal
  cluster randomised trials.
\newblock {\em Statistics in Medicine}, 35(26):4718--4728, jun 2016.

\bibitem{Reich2012}
Nicholas~G. Reich, Jessica~A. Myers, Daniel Obeng, Aaron~M. Milstone, and
  Trish~M. Perl.
\newblock Empirical power and sample size calculations for cluster-randomized
  and cluster-randomized crossover studies.
\newblock {\em PLOS ONE}, 7(4):1--7, 04 2012.

\bibitem{Wilson2015}
D.~T. Wilson, R.~E. Walwyn, J.~Brown, A.~J. Farrin, and S.~R. Brown.
\newblock Statistical challenges in assessing potential efficacy of complex
  interventions in pilot or feasibility studies.
\newblock {\em Statistical Methods in Medical Research}, 25(3):997--1009, jun
  2015.

\bibitem{Deb2002}
K.~Deb, A.~Pratap, S.~Agarwal, and T.~Meyarivan.
\newblock A fast and elitist multiobjective genetic algorithm: {NSGA}-{II}.
\newblock {\em {IEEE} Transactions on Evolutionary Computation}, 6(2):182--197,
  apr 2002.

\bibitem{Mersmann2014}
Olaf Mersmann.
\newblock {\em mco: Multiple Criteria Optimization Algorithms and Related
  Functions}, 2014.
\newblock R package version 1.0-15.1.

\bibitem{Sacks1989}
Jerome Sacks, William~J. Welch, Toby~J. Mitchell, and Henry~P. Wynn.
\newblock Design and analysis of computer experiments.
\newblock {\em Statistical Science}, 4(4):409--423, 1989.

\bibitem{Santner2003}
Thomas~J. Santner, Brian~J. Williams, and William~I. Notz.
\newblock {\em The Design and Analysis of Computer Experiments}.
\newblock Springer-Verlag New York, Inc., 2003.

\bibitem{Krige1951}
D.G. Krige.
\newblock A statistical approach to some basic mine valuation problems on the
  witwatersrand.
\newblock {\em Journal of the Southern African Institute of Mining and
  Metallurgy}, 52(6):119--139, 1951.

\bibitem{Rasmussen2006}
Carl~Edward Rasmussen and Christopher K.~I. Williams.
\newblock {\em Gaussian Processes for Machine Learning}.
\newblock MIT Press, 2006.

\bibitem{Roustant2012}
Olivier Roustant, David Ginsbourger, and Yves Deville.
\newblock {DiceKriging}, {DiceOptim}: Two {R} packages for the analysis of
  computer experiments by kriging-based metamodeling and optimization.
\newblock {\em Journal of Statistical Software}, 51(1):1--55, 2012.

\bibitem{Jones2001}
Donald~R. Jones.
\newblock A taxonomy of global optimization methods based on response surfaces.
\newblock {\em Journal of Global Optimization}, 21(4):345--383, 2001.

\bibitem{Sasena2002}
Michael~J. Sasena, Panos Papalambros, and Pierre Goovaerts.
\newblock Exploration of metamodeling sampling criteria for constrained global
  optimization.
\newblock {\em Engineering Optimization}, 34(3):263--278, jan 2002.

\bibitem{Emmerich2011}
Michael~TM Emmerich, Andr{\'e}~H Deutz, and Jan~Willem Klinkenberg.
\newblock Hypervolume-based expected improvement: Monotonicity properties and
  exact computation.
\newblock In {\em Evolutionary Computation (CEC), 2011 IEEE Congress on}, pages
  2147--2154. IEEE, 2011.

\bibitem{Picheny2014}
Victor Picheny and David Ginsbourger.
\newblock Noisy kriging-based optimization methods: A unified implementation
  within the {DiceOptim} package.
\newblock {\em Computational Statistics {\&} Data Analysis}, 71:1035--1053, mar
  2014.

\bibitem{Browne2009}
William~J Browne, Mousa~Golalizadeh Lahi, and Richard~MA Parker.
\newblock {\em A Guide to Sample Size Calculations for Random Effect Models via
  Simulation and the MLPowSim Software Package}, 2009.

\bibitem{White2007}
Peter White, Michael Sharpe, Trudie Chalder, Julia DeCesare, Rebecca Walwyn,
  and the PACE~trial group.
\newblock Protocol for the pace trial: A randomised controlled trial of
  adaptive pacing, cognitive behaviour therapy, and graded exercise as
  supplements to standardised specialist medical care versus standardised
  specialist medical care alone for patients with the chronic fatigue
  syndrome/myalgic encephalomyelitis or encephalopathy.
\newblock {\em BMC Neurology}, 7(1):6, 2007.

\bibitem{White2011}
PD~White, KA~Goldsmith, AL~Johnson, L~Potts, R~Walwyn, JC~DeCesare, HL~Baber,
  M~Burgess, LV~Clark, DL~Cox, J~Bavinton, BJ~Angus, G~Murphy, M~Murphy,
  H~O'Dowd, D~Wilks, P~McCrone, T~Chalder, and M~Sharpe.
\newblock Comparison of adaptive pacing therapy, cognitive behaviour therapy,
  graded exercise therapy, and specialist medical care for chronic fatigue
  syndrome (pace): a randomised trial.
\newblock {\em The Lancet}, 377(9768):823--836, 2011.

\bibitem{Chalder1993}
T~Chalder, G~Berelowitz, S~Hirsch, T~Pawlikowska, P~Wallace, and S~Wessely.
\newblock Development of a fatigue scale.
\newblock {\em Journal of Psychometric Research}, 37(2):147--153, 1993.

\bibitem{Walwyn2010}
Rebecca E.~A. Walwyn and Chris Roberts.
\newblock Therapist variation within randomised trials of psychotherapy:
  implications for precision, internal and external validity.
\newblock {\em Statistical Methods in Medical Research}, 19(3):291--315, 2010.

\bibitem{Roberts2015}
Chris Roberts, Evridiki Batistatou, and Stephen~A. Roberts.
\newblock Design and analysis of trials with a partially nested design and a
  binary outcome measure.
\newblock {\em Statistics in Medicine}, 35(10):1616--1636, 2015.

\bibitem{McHorney1993}
Colleen~A. McHorney, John~E. Ware, and Anastasia~E. Raczek.
\newblock The mos 36-item short-form health survey (sf-36): Ii. psychometric
  and clinical tests of validity in measuring physical and mental health
  constructs.
\newblock {\em Medical Care}, 31(3):247--263, 1993.

\bibitem{Senn2007}
Stephen Senn and Frank Bretz.
\newblock Power and sample size when multiple endpoints are considered.
\newblock {\em Pharmaceut. Statist.}, 6(3):161--170, 2007.

\bibitem{Jones1998}
Donald~R. Jones, Matthias Schonlau, and William~J. Welch.
\newblock Efficient global optimization of expensive black-box functions.
\newblock {\em Journal of Global Optimization}, 13(4):455--492, 1998.

\bibitem{Bendtsen2012}
Claus Bendtsen.
\newblock {\em pso: Particle Swarm Optimization}, 2012.
\newblock R package version 1.0.3.

\bibitem{Picheny2010}
V.~Picheny, D.~Ginsbourger, and Y.~Richet.
\newblock Noisy expected improvement and on-line computation time allocation
  for the optimization of simulators with tunable fidelity.
\newblock In {\em 2nd International Conference on Engineering Optimization},
  2010.

\bibitem{Dutang2015}
Christophe Dutang and Petr Savicky.
\newblock {\em randtoolbox: Generating and Testing Random Numbers}, 2015.
\newblock R package version 1.17.

\bibitem{Chevalier2014}
Cl{\'{e}}ment Chevalier, Victor Picheny, and David Ginsbourger.
\newblock {KrigInv}: An efficient and user-friendly implementation of
  batch-sequential inversion strategies based on kriging.
\newblock {\em Computational Statistics {\&} Data Analysis}, 71:1021--1034, mar
  2014.

\bibitem{Bates2015}
Douglas Bates, Martin M{\"a}chler, Ben Bolker, and Steve Walker.
\newblock Fitting linear mixed-effects models using {lme4}.
\newblock {\em Journal of Statistical Software}, 67(1):1--48, 2015.

\bibitem{Goldstein2003}
Harvey Goldstein.
\newblock {\em Multilevel Statistical Models}.
\newblock Arnold, 3rd edition, 2003.

\bibitem{Hemming2017}
K~Hemming, S~Eldridge, G~Forbes, C~Weijer, and M~Taljaard.
\newblock How to design efficient cluster randomised trials.
\newblock {\em BMJ}, 358, 2017.

\bibitem{Donner2000}
Allan Donner and Neil Klar.
\newblock {\em Design and Analysis of Cluster Randomization Trials in Health
  Research}.
\newblock London Arnold Publishers, 2000.

\bibitem{Hemming2011}
Karla Hemming, Alan Girling, Alice Sitch, Jennifer Marsh, and Richard Lilford.
\newblock Sample size calculations for cluster randomised controlled trials
  with a fixed number of clusters.
\newblock {\em BMC Medical Research Methodology}, 11(1):102, 2011.

\bibitem{Eldridge2015}
Sandra~M Eldridge, Ceire~E Costelloe, Brennan~C Kahan, Gillian~A Lancaster, and
  Sally~M Kerry.
\newblock How big should the pilot study for my cluster randomised trial be?
\newblock {\em Statistical Methods in Medical Research}, 2015.

\bibitem{Hox2002}
Joop Hox.
\newblock {\em Multilevel Analysis: Techniques and Applications}.
\newblock Lawrence Erlbaum Associates, Inc., 2002.

\bibitem{Snijders1993}
Tom A.~B. Snijders and Roel~J. Bosker.
\newblock Standard errors and sample sizes for two-level research.
\newblock {\em Journal of Educational and Behavioral Statistics},
  18(3):237--259, 1993.

\bibitem{Raudenbush2000}
Stephen~W. Raudenbush and Xiaofeng Liu.
\newblock Statistical power and optimal design for multisite randomized trials.
\newblock {\em Psychological Methods}, 5(2):199--213, 2000.

\bibitem{Breukelen2012}
Gerard~J.P. van Breukelen and Math~J.J.M. Candel.
\newblock Calculating sample sizes for cluster randomized trials: We can keep
  it simple and efficient!
\newblock {\em Journal of Clinical Epidemiology}, 65(11):1212 -- 1218, 2012.

\bibitem{Teerenstra2008}
S.~Teerenstra, M.~Moerbeek, T.~van Achterberg, B.~J. Pelzer, and G.~F. Borm.
\newblock Sample size calculations for 3-level cluster randomized trials.
\newblock {\em Clinical Trials}, 5(5):486--495, sep 2008.

\bibitem{Joseph1997a}
Lawrence Joseph and David~B. Wolfson.
\newblock Interval-based versus decision theoretic criteria for the choice of
  sample size.
\newblock {\em Journal of the Royal Statistical Society: Series D (The
  Statistician)}, 46(2):145--149, 1997.

\bibitem{Smith2010}
Mike~K. Smith and Andrea Marshall.
\newblock Importance of protocols for simulation studies in clinical drug
  development.
\newblock {\em Statistical Methods in Medical Research}, 2010.

\bibitem{Burton2006}
Andrea Burton, Douglas~G. Altman, Patrick Royston, and Roger~L. Holder.
\newblock The design of simulation studies in medical statistics.
\newblock {\em Statistics in Medicine}, 25(24):4279--4292, 2006.

\bibitem{Kennedy2001}
Marc~C. Kennedy and Anthony O'Hagan.
\newblock Bayesian calibration of computer models.
\newblock {\em Journal of the Royal Statistical Society: Series B (Statistical
  Methodology)}, 63(3):425--464, 2001.

\bibitem{OHagan2005}
Anthony O'Hagan, John~W. Stevens, and Michael~J. Campbell.
\newblock Assurance in clinical trial design.
\newblock {\em Pharmaceutical Statistics}, 4(3):187--201, 2005.

\bibitem{Cao2009}
Jing Cao, J.~Jack Lee, and Susan Alber.
\newblock Comparison of bayesian sample size criteria: {ACC}, {ALC}, and {WOC}.
\newblock {\em Journal of Statistical Planning and Inference}, 139(12):4111 --
  4122, 2009.

\bibitem{Oakley2010}
Jeremy~E. Oakley, Alan Brennan, Paul Tappenden, and Jim Chilcott.
\newblock Simulation sample sizes for monte carlo partial evpi calculations.
\newblock {\em Journal of Health Economics}, 29(3):468 -- 477, 2010.

\bibitem{Wason2012}
James M.~S. Wason and Thomas Jaki.
\newblock Optimal design of multi-arm multi-stage trials.
\newblock {\em Statistics in Medicine}, 31(30):4269--4279, jul 2012.

\end{thebibliography}

\end{document}